\documentclass[aps,prc,nofootinbib,showpacs]{revtex4}

\usepackage{epsfig}
\usepackage{graphicx}

\begin{document}
\title{$P$-parity of charm and strange particles in electroproduction processes, in collinear regime.}
\author{Michail P. Rekalo }
\affiliation{National Science Centre - Kharkov Institute of 
Physics and Technology,\\ Akademicheskaya 1, 61108 Kharkov,
Ukraine}
\author{Egle Tomasi-Gustafsson}
\affiliation{\it DAPNIA/SPhN, CEA/Saclay, 91191 Gif-sur-Yvette Cedex, 
France}
\date{\today}


\begin{abstract}
We show that definite polarization observables for the simplest electroproduction processes, $\ell+N\to \ell+B(1/2^\pm)+M(0^\pm)$, $B=Y(\Lambda, \Sigma$ or $Y_c$-hyperon, $\Theta^+$-pentaquark) and $M=K$, $\overline{K}$ or $D$, are sensitive to the relative P-parity $\pi(B)$ of the NBM-system. The interference of the longitudinal and transversal amplitudes for the collinear regime of the processes $\gamma^*+N\to B+M $($\gamma^*$ is the virtual photon) - at any value of momentum transfer squared and excitation energy of the $BM$-system - generates model independent relations between analyzing powers ( in unpolarized lepton scattering by polarized target), from one side, and the components of the produced baryon $B$ polarization. It is important to stress that these relations depend on the above mentioned P-parity, and constitute a model independent method for the determination of unknown parities of strange and charm particles.

\end{abstract}
\maketitle

Parity is a fundamental property of elementary particles, therefore its experimental determination is very important. Sometimes this is a very difficult problem, as in case of $\Theta^+$-pentaquark \cite{Di97}, which has been recently observed in different experiments \cite{Na03}. However, these experiments show the existence of a narrow structure in a missing mass spectrum, and, at best, can determine the mass and the with of the resonance and can not determine unambiguously the spin and parity of the $\Theta^+$-hyperon. Therefore, this problem is very actual, because the complete understanding of the structure of $\Theta^+$ depends essentially on the parity of this state. Theoretical predictions, \cite{Cs03} allow both values of the $\Theta^+$-parity. Note, in this respect, that the P-parity, which is conserved in strong and electromagnetic interaction,  cannot be determined for the $\Theta^+$, as an absolute quantum number (equal to $\pm 1$), as in case of neutral particles as $\gamma$, $\pi^0$, $\eta$, $\rho$ etc., because the strangeness of the  $\Theta^+$ is nonzero. Only the relative P-parity  of the system $P(N\Theta K)$ has a physical sense, because the decay $\Theta^+\to N+K$ is the main decay, and all dynamics of such processes as $\gamma+N\to \Theta^+ +\overline{K}$,  $\pi+N\to \Theta^+ +\overline{K}$, $N+N\to \Sigma +\Theta^+ 
$, will depend namely on  $P(N\Theta K)$. Recent calculations for  $\gamma+N\to \Theta^+ +\overline{K}$ \cite{Zh04} showed the sensitivity of some polarization observables on $P(N\Theta K)$. Unfortunately, most of  these predictions are model dependent, and the necessity of model independent guides for the determination of the $\Theta^+$-parity, are, in our opinion, unavoidable. Model independent methods are based on the study of polarization observables. Such methods have been firstly suggested for the determination of the (relative) parity of strange particles in $\pi^-+p\to \Lambda^0 +K^0$, comparing the signs of the analyzing power and of the $\Lambda$ polarization \cite{Bi58}. Later, it 
was suggested \cite{Pa99} that polarization phenomena in $p+p\to \Lambda^0+K^++p$ can be also very useful, for this aim.

These methods are based on general properties of fundamental interactions, therefore they can also be applied to charm particles (for example, $\pi+N\to \Lambda_c^++\overline{D}$, $p+p\to p+\Lambda_c^+ +\overline{D}^0$) as well. It is straightforward to adapt such methods to the determination of the P-parity of the $\Theta^+ $ pentaquark: $\pi^-+p\to \Theta^+ +K^-$ \cite{Re04a}, 
$ p+p\to \Sigma^+ +\Theta^+ $\cite{Th03,Re04b}, $ n+p\to \Lambda^0
+\Theta^+ $\cite{Re04a,Re04b,Uz04}, $ p+p\to \pi^++\Lambda^0+p$ \cite{Re04a}, and 
$\gamma+p\to \Theta^+ +\overline{K^0}$ \cite{Re04c,Na04}.
In the last case, the determination of the $\Theta^+ $ P-parity can be done in different ways, as several relations among polarization observables depend on the relative $N\Theta K$-parity.

Our aim, here, is to consider the simplest process of associative electroproduction:
$$e^-+N\to e^- +B(1/2^\pm)+M(0^\pm),$$
where $B(1/2^\pm)$ is a baryon with spin $J$ and parity P, $J^P=1/2^{\pm}$ and $M(0^{\pm})$ is a scalar or pseudoscalar meson, as a possible source of information about the relative $NBM$-parity. The present study will be done in a model independent form, and, therefore, the results can be applied to different physical cases, i.e. our considerations hold for $Y+K$ ($Y=\Lambda$ or $\Sigma)$, $\Lambda_c^+ +\overline{D}$ and $\Theta^+ +K$-production. To simplify the analysis, we consider the collinear regime for 
$\gamma^*+N\to  B(1/2^\pm)+M(0^\pm)$, $\gamma^*$ is the virtual photon, but the results hold for any value of the space-like momentum transfer $Q^2$ and for any value of the effective mass $W$ of the produced $BM$ system. The advantages of the collinear regime are an essential simplification of the spin structure of the corresponding matrix element, and a corresponding simplification of the analysis of polarization phenomena, with a small number of kinematical variables. In the collinear regime, generally the cross section takes its maximal value, in particular at large $W$. 


The spin structure of the  matrix element for $\gamma^*+N\to B(1/2^\pm) +M(0^\pm)$, in collinear kinematics, in the center of mass (c.m.) of the considered reaction,  can be parametrized in the following general form (using the P-invariance of the hadronic electromagnetic interaction and the conservation of the electromagnetic current):
$$
{\cal M}^{(\pm)}=\chi^{\dagger}_2{\cal F}^{(\pm)} {\chi}_1,
$$
\begin{equation}
{\cal F}^{(+)}=\vec\sigma\cdot\vec e\times\hat{\vec k} f_t^{(+)}(Q^2,W)+i\vec e\cdot\hat{\vec k}f_{\ell}^{(+)}(Q^2,W),~\mbox{~if~} \pi(B)=P(NBM)=+ 1, 
\label{eq:mat}
\end{equation}
\begin{equation}
{\cal F}^{(-)}=\vec\sigma\cdot\vec e  f_t^{(-)}(Q^2,W)+\vec e 
\cdot\hat{\vec k}\vec\sigma\cdot\hat{\vec k}f_{\ell}^{(-)}(Q^2,W)
,~\mbox{~if~} \pi(B)=P(NBM)=- 1, 
\label{eq:mat1}
\end{equation}
where $\vec e$ is the three vector of the virtual photon polarization , $\chi_1$ and $\chi_2$  are the
two-component spinors of the initial nucleon and the final baryon, 
$\hat{\vec k}$ is the unit vectors along the three momentum of the virtual photon, $f_{\ell,t}^{(\pm)}$ are the collinear amplitudes, which are, generally, complex functions of $Q^2$ and $W$, and describe the absorption of a $\gamma^*$ with transversal $(t)$ or longitudinal $(\ell)$ polarization. The upper index, $(\pm)$, corresponds to two different possible values of the relative $\pi$-parity.

Such general form for the collinear amplitudes, Eqs. (\ref{eq:mat}) and (\ref{eq:mat1}), allows us to calculate any polarization observables for the process $\gamma^*+N\to B+M $.

The dependence of the differential cross section for $\vec N(e,e'B)M$ (or $\vec N,(e,e'M)B$) on the target polarization, $\vec P$, is characterized by the following tensor:
\begin{equation}
{\cal A}^{(\pm)}_{mn}=\displaystyle\frac{1}{2}Tr {\cal F}_m^{(\pm)} \vec\sigma\cdot\vec P {\cal F}_n^{(\pm)\dagger}, 
\label{eq:asy}
\end{equation}
where ${\cal F}^{(\pm)}=e_m{\cal F}^{(\pm)}_m$. Note that Eq. (\ref{eq:asy}) corresponds to the case of the production of unpolarized baryon $B$.

Applying the P-invariance of electromagnetic interaction (of strange and charm particles), the tensor ${\cal A}^{(\pm)}_{mn}$ can be written in the following general form, where the dependence of the target polarization $\vec P$ appears explicitly:
\begin{equation}
{\cal A}^{(\pm)}_{mn}=i\epsilon_{mn\ell}P_{\ell} A_1^{(\pm)}(Q^2,W) +\epsilon_{mn\ell}\hat k_{\ell}(\hat{\vec k}\cdot \vec P) 
A_2^{(\pm)}(Q^2,W)+[\hat k_m(\hat{\vec k}\times \vec P)_n+\hat k_n( \hat{\vec k}\times \vec P)_m]A_3^{(\pm)}(Q^2,W),
\label{eq:aspm}
\end{equation}
where $A_i^{(\pm)}(Q^2,W)$, $i=1-3$, are the polarized structure functions, (SFs), being real functions of $Q^2,W$.

The symmetrical part of the tensor $ {\cal A}^{(\pm)}_{mn}$ determine the scattering of unpolarized leptons by polarized target (in direction perpendicular to the photon three momentum). The corresponding SF, $A_3^{(\pm)}(Q^2,W)$, being T-odd, is determined by the interference of the longitudinal and transversal collinear amplitudes, $f_{\ell}^{(\pm)}$ and $f_{t}^{(\pm)}$ - for any P-parity. 

The antisymmetrical part of $ {\cal A}^{(\pm)}_{mn}$, with the SFs $A_{1,2}^{(\pm)}(Q^2,W)$, determines the scattering of longitudinally polarized leptons.

Using Eqs. (\ref{eq:mat}) and (\ref{eq:mat1}), one can find the following formulas for the SFs $A_{12}^{(\pm)}(Q^2,W)$, in terms of collinear amplitudes:
$$
A_1^{(\pm)}(Q^2,W)=-A_2^{(\pm)}(Q^2,W)=Re f_t^{(\pm)}(Q^2,W)f_{\ell}^{(\pm)*}(Q^2,W),
$$
\begin{equation}
A_3^{(\pm)}(Q^2,W)=Imf_t^{(\pm)}(Q^2,W)f_{\ell}^{(\pm)*}(Q^2,W),
\label{eq:as123}
\end{equation}
i.e. both value of $P(NBM)$ give rise to identical formulas. This indicates that polarization observables, which arise from the target polarization only, can not discriminate, in model independent way, the parity $P(NBM)$.


The polarization of the produced baryon is characterized by the following construction:
\begin{equation}
{\cal P}^{(\pm)}_{mn,\ell}=\displaystyle\frac{1}{2}Tr {\cal F}^{(\pm)}_m {\cal F}^{(\pm)\dagger}_n 
\vec\sigma\cdot\vec \ell, 
\label{eq:pol}
\end{equation}
where $\vec\ell$ is the unit vector, determining the direction of the the final baryon polarization.

Again, the P-invariance of electromagnetic interaction allows us to parametrize ${\cal P}^{(\pm)}_{mn,\ell}$ in a general, model independent form:
\begin{equation}
{\cal P}^{(\pm)}_{mn,\ell}=i\epsilon_{mn\ell}P_1^{(\pm)}(Q^2,W) +i\epsilon_{mna}\hat k_a\hat k_{\ell}
P_2^{(\pm)}(Q^2,W)+
[\hat{k}_m \epsilon_{na\ell}\hat k _a+  
\hat k_n\epsilon_{ma\ell}\hat k_a]P_3^{(\pm)}(Q^2,W),
\label{eq:polcm}
\end{equation}
where $P_i^{(\pm)}(Q^2,W)$, $i=1-3$, are the real SFs, characterizing the polarization properties of $B$ (for collinear electroproduction), and they are quadratic combinations of the collinear amplitudes.

Using Eqs. (\ref{eq:mat}), (\ref{eq:mat1}) and (\ref{eq:as123}) one can find:
\begin{eqnarray}
P_{1,2}^{(\pm)}(Q^2,W)&=&-A_{1,2}^{(\pm)}(Q^2,W),\nonumber\\
P_3^{(\pm)}(Q^2,W)&=&-\pi(B) A_3^{(\pm)}(Q^2,W).
\label{eq:pol12}
\end{eqnarray}
The relation (\ref{eq:pol12}) is the main result of this paper, which can be formulated in the following general form: in the kinematical conditions of collinear electroproduction, for $e^-+N\to e^-+ B+M$ (where the final hadron is emitted in the direction of the photon three-momentum) two polarization observables, 
$P_3^{(\pm)}(Q^2,W)$ and $ A_3^{(\pm)}(Q^2,W)$ are sensitive to the relative $P(NBM)$-parity, at any value of  $Q ^2$, and $W$. It is a model independent, general result.

We refer here to the simplest case of unpolarized lepton scattering, therefore the discussed polarization observables should be proportional to the product $\ell_{mn}{\cal A}^{(\pm)}_{mn}$ or $\ell_{mn}{\cal P}^{(\pm)}_{mn,\ell}$, with the following expression for the leptonic tensor:
\begin{equation}
\ell_{mn}\simeq k_{1m}k_{2n}+k_{1n}k_{2m}+\delta_{mn}Q^2,
\label{eq:all}
\end{equation}
where $\vec k_1$ and $\vec k_2$ are the three momenta of the initial and final electrons. Comparing Eqs. (\ref{eq:aspm}) and (\ref{eq:all}), one can see that the SF $A_3^{(\pm)}(Q^2,W)$ determines the asymmetry in $e^-+\vec N\to e^-+ B+M$, induced by the target polarization which has to be orthogonal to the electron scattering plane, and the SF $P_3^{(\pm)}(Q^2,W)$ determines the component of the B-polarization, in the same direction.


In conclusion, we derived in model independent way, a relation among polarization observables in the reaction $e^-+ N\to e^-+ B+M$, Eq. (\ref{eq:pol12}), which is sensitive to the relative P-parity $P(NBM)$. This result can be considered a generalization of a corresponding relation between the analyzing power in $\pi^-+\vec p\to \Lambda^0+K^0$ and the transversal $\Lambda$ polarization in  $\pi^-+\vec p\to \vec\Lambda^0+K^0$, which has been suggested many years ago in \cite{Bi58} as a model independent way for the determination of the parity of strange particles.
 
In this respect, the following important dynamical ingredients can be noted here:
\begin{itemize}
\item the discussed single-spin polarization observables, for $e^-+ N\to e^-+ B+M$ are determined by the interference of transversal and longitudinal collinear amplitudes, $f_{1,2}^{(\pm)*}(Q^2,W)$. Therefore, the suggested method holds only in the case of electroproduction, i.e. with virtual photons. There is no similar possibility in photoproduction processes, where the  photon polarization is only transversal. We proved earlier \cite{Re04c}, that only particular triple-spin polarization observables for $\gamma+N\to B+M $, in collinear regime, are sensitive to the relative P-parity. The same is correct also for collinear electroproduction \cite{Re03}, taking into account the contributions of transversal or longitudinl virtual photon, separately.
\item The discussed polarization observables for $e+N\to e+B+M$ (the SFs $A_3^{(\pm)}(Q^2,W)$ and $P_3^{(\pm)}(Q^2,W)$), being T-odd, are proportional to $ Im f_t f_{\ell}^*$, and are nonzero for the complex collinear amplitudes, with different phases $\delta_{\ell}^{(\pm)}-\delta_t^{(\pm)}\ne 0$. The exact
value of these phases do not affect the realization of the suggested method.
\end{itemize}
Note, in this respect, that existing models for the electroproduction of $YK$ or $\Theta K$-system generate a large difference  $\delta_{\ell} ^{(\pm)}-\delta_t^{(\pm)}$, i.e. large absolute values for the discussed T-odd polarization observables - independently on the kinematical region of the variables $Q^2$ and $W$. Indeed, in the resonance region (corresponding to $W\simeq$ 2$\div$ 3 GeV, and $Q^2$ any, in the space-like region) the contribution of several nucleonic resonances should also be taken into account - with resulting essential complexity of the collinear amplitudes, transversal and longitudinal.

At higher values of excitation energy, where the Regge pole approach applies, the possible Regge $K$ and $K^+$ exchange (with different trajectories) in forward direction, or fermionic ($Y$, $Y'$)-exchange in backward direction are also source of essential T-odd effects.

The situation for charm electroproduction is similar - with respect to the possible size of T-odd effects. Of course, due to the higher threshold of $Y_c \overline{D}$-production, the effect of possible nucleonic resonances should be negligible. But, in the Regge description, at higher $W$, the reggeized $D$ and $D^*$ exchange contributions, again with different trajectories, can generate naturally complex amplitudes with different phases.

Note that the model independent method suggested here, for the determination of $P(NBM)$, implies the measurement of the polarization of the emitted baryon. Experimentally, it is a difficult measurement for the $\Theta^+$ baryon, as it requires the measurement of the proton polarization in the strong decay $\Theta^+\to p+K^0$. In this respect, the situation with  $Y$ and $Y_c$ baryons is easier, because all these particles, decaying through the weak interaction, are self analyzing particles.

{}

\end{document}